\documentclass[]{imsart}

\usepackage{amsthm,amsmath}
\RequirePackage[numbers]{natbib}
\usepackage{graphicx}
\RequirePackage[colorlinks,citecolor=blue,urlcolor=blue]{hyperref}
\usepackage{bbold}

\startlocaldefs
\newcommand{\R}{\mathbb{R}}

\numberwithin{equation}{section}
\newtheorem{theorem}{Theorem}[section]
\endlocaldefs

\begin{document}

\begin{frontmatter}

\title{ROC Analysis for Paired Comparison Data}
\runtitle{ROC Analysis for Paired Comparison Data}


\author{\fnms{Ran} \snm{Huo}\thanksref{t1}\corref{}\ead[label=e1]{rhuo@g.harvard.edu}}
\and
\author{\fnms{Mark E.} \snm{Glickman}\ead[label=e2]{glickman@fas.harvard.edu}}
\thankstext{t1}{To whom correspondence should be addressed.}
\address{Department of Statistics\\
Harvard University\\
1 Oxford Street, Cambridge, MA 02138, USA\\
\printead{e1,e2}
}

\runauthor{R. Huo and M. E. Glickman}

\begin{abstract}
Paired comparison models are used for analyzing data that involves pairwise comparisons among a set of objects. When the outcomes of the pairwise comparisons have no ties, the paired comparison models can be generalized as a class of binary response models. Receiver operating characteristic (ROC) curves and their corresponding areas under the curves are commonly used as performance metrics to evaluate the discriminating ability of binary response models. Despite their individual wide range of usage and their close connection to binary response models, ROC analysis to our knowledge has never been extended to paired comparison models since the problem of using different objects as the reference in paired comparison models prevents traditional ROC approach from generating unambiguous and interpretable curves. We focus on addressing this problem by proposing two novel methods to construct ROC curves for paired comparison data which provide interpretable statistics and maintain desired asymptotic properties. The methods are then applied and analyzed on head-to-head professional sports competition data. 
\end{abstract}




\end{frontmatter}

\section{Introduction}
\label{sec:intro}
Paired comparison data arise from situations where a set of objects are being compared in pairs. Various models for analyzing paired comparison data have been proposed and developed in the past few decades, among which the most widely used ones are the Bradley-Terry model \citep{bradley1952rank} and the Thurstone-Mosteller model 
\citep{thurstone1927law,mosteller1951remarks}. \citet{davidson1976bibliography} provided a comprehensive bibliography of paired comparison literature in various areas of study, including statistics, psychometrics, marketing research, preference measurement, and sports.  A more recent review of the extensions developed for Bradley-Terry and Thurstone-Mosteller models was given by \citet{cattelan2012models}. 

The Bradley-Terry and Thurstone-Mosteller models are examples of ``linear'' paired comparison models \citep{DavidH.A.HerbertAron1988Tmop} insofar as the probability of a preferred outcome depends on the difference between object-specific parameters.  These two models in fact are special cases of logistic regression and probit regression for a binary response, respectively, as illustrated by \cite{critchlow1991paired}.
One of the commonly used techniques for assessing the performance of binary response models is by constructing the ROC curve. The area under the ROC curve, which is referred to as the AUC or the c-statistic, evaluates the ability of binary models at discriminating between positive and negative observations and is commonly used as a performance metric. However, the traditional ROC analysis fails to produce reliable results for paired comparisons since the interpretations of ROC curves rely on clearly defining ``success'' and ``failure'' for a binary outcome, while for paired comparison data, they are defined ambiguously. Therefore, in this paper, we focus on resolving this ambiguity and propose two approaches to extend ROC analysis for paired comparison data. We also apply the two approaches to the games played in the 2017 or the 2017-18 regular-season Major League Baseball (MLB), National Hockey League (NHL), National Football League (NFL), and National Basketball League (NBA), and use the corresponding AUC statistics to compare the level of parity for these professional sports \citep{fort1995cross}.

In Section~\ref{sec:models}, we give a brief review of paired comparison models and introduce the assumptions that we make and the notations that we use for the remaining of the paper. In Section~\ref{sec:ROC}, we first identify the problems of applying traditional ROC analysis to paired comparison data. Then we describe two methods of extending the traditional approach to construct ROC curves for paired comparison data. We also redefine the c-statistics associated with the corresponding ROC curves and discuss their asymptotic properties in this section. Section~\ref{sec:comparison} discusses the relationship between the two redefined c-statistics and compares their standard errors. In Section~\ref{sec:applications}, we present and analyze the results of applying the two approaches to the professional sports data set.  We conclude our paper in Section~\ref{sec:conclusions}.

\section{Linear paired comparison models}
\label{sec:models}

Consider a set of comparisons involving $m$ teams, players, treatments, or objects. We will henceforth refer to the objects being compared as \textit{teams} in the context of head-to-head sports. For $i=1,\dots,m$, assume a team strength parameter $\mu_i\in\R$ that measures the true ability of team $i$. Let $Y_{ij}$ be a binary random variable indicating the result of competition between team $i$, and team $j$, such that $Y_{ij}=1$ if team $i$ wins and $Y_{ij}=0$ if team $i$ loses. We assume no ties or partial preferences. Let $p_{ij}$ denote the probability of team $i$ defeating team $j$, and assume $Y_{ij}\sim\text{Bernoulli}(p_{ij})$.

A linear paired comparison model assumes that 
\begin{equation}
\label{equa:pcm}
    p_{ij}=F(\mu_i-\mu_j),
\end{equation} 
where $F$ is the cumulative distribution function (CDF) of a continuous probability distribution \citep{DavidH.A.HerbertAron1988Tmop}. Notice that $Y_{ij}=1$ is equivalent to $Y_{ji}=0$ since they both indicate team $i$ winning. Thus, 
\begin{equation*}
    F(\mu_i-\mu_j)=p_{ij}=1-p_{ji}=1-F(\mu_j-\mu_i),
\end{equation*}
which requires $F$ to be the CDF of a symmetric distribution.

The model in (~\ref{equa:pcm}) includes several popular special cases.
When $F$ is specified to be a standard logistic CDF, the model can be written as
\begin{equation}
\label{equa:bradley-terry}
    p_{ij}=\frac{1}{1+e^{-(\mu_i-\mu_j)}},
\end{equation}
or equivalently,
\begin{equation*}
    p_{ij}=\frac{e^{\mu_i}}{e^{\mu_i}+e^{\mu_j}}.
\end{equation*}
This model is known as the Bradley-Terry model \citep{bradley1952rank}. 
When $F$ is the standard normal CDF, the model has the form
\begin{equation*}
    p_{ij}=\Phi(\mu_i-\mu_j),
\end{equation*}
which is called the Thurstone-Mosteller model 
\citep{thurstone1927law,mosteller1951remarks}. 

Inference for
linear paired comparison models can be obtained through maximum likelihood estimation (MLE). 
Letting $\boldsymbol{\hat{\mu}}=(\hat{\mu}_1,\hat{\mu}_2,\dots,\hat{\mu}_m)$ 
be the MLEs of the team strength parameters $\boldsymbol{\mu}=(\mu_1,\mu_2,\dots,\mu_m)$, 
it follows that an estimate of $p_{ij}$ can be calculated as 
$\hat{p}_{ij}=F(\hat{\mu}_i-\hat{\mu}_j)$ by replacing the $\mu_i$ with the MLEs in (\ref{equa:pcm}). 
To ensure identifiability of the models, a linear constraint is usually assumed on the strength parameters; for example, it is commonly assumed that $\sum_{i=1}^{m}\mu_{i}=0$, or $\mu_i=0$ for some $i\in\{1,\dots,m\}$. In addition, we assume that for every possible partition of the teams into two nonempty sets, some team in the second set has defeated a team in the first set at least once. This partition assumption by \citet{ford1957solution} guarantees the existence of a unique MLE within the constraint region. 
Letting each team be a node in a graph, and letting a directed edge from node $i$ to node $j$ represent a win by $i$ over $j$, this assumption is equivalent to  there always exists a path from $i$ to $j$  for all $i$ and $j$ \citep{hunter2004mm}. 

For the remainder of the development, let $[n]$ denote the set $\{1,\ldots, n\}$, and let $|S|$ denote the number of elements in a set $S$. Let $R=\binom{m}{2}$ be the number of all possible pairs among $m$ teams. All further notation involving $i$ and $j$ are defined for $i,j\in[m]$ and $i<j$. Let $n_{ij}$ be the total number of comparisons between team $i$ and team $j$, and then the total number of paired comparisons $N$ can be defined as $N=\sum_{i,j\in[m],i<j}n_{ij}$. Let $w_{ij}$ be the number of times team $i$ defeats team $j$. For $k\in [n_{ij}]$, where $n_{ij}>0$, let $Y_{ijk}$ represent the binary outcome of the $k^{th}$ competition between team $i$ and team $j$, then $w_{ij}=\sum_{k=1}^{n_{ij}}Y_{ijk}$.
We assume that
all competitions are conditionally independent;
that is, $Y_{ijk}|p_{ij}\overset{\text{i.i.d.}}{\sim}\text{Bern}(p_{ij})$.

\section{ROC analysis}
\label{sec:ROC}
In this section, we first review ROC curves for binary response models. We then focus on extensions of ROC analyses for paired comparison data and propose two approaches. We conclude this section with a discussion of the asymptotic behavior of ROC curves constructed from those two approaches.

\subsection{ROC curves for binary response models}
\label{subsec:ROCbin}
ROC curves are often used to analyze the discriminatory performance of binary response models. Suppose $Y_i$ is a binary response defined as
\begin{equation*}
Y_{i}=\left\{
  \begin{array}{@{}ll@{}}
    0,&  \text{if $i$ is a ``failure",} \\
    1,&  \text{if $i$ is a ``success".}\\
  \end{array}\right.
\end{equation*}
Letting $p_i$ denote the probability of success, we assume $Y_i\sim\text{Bern}(p_i)$. Let
$\hat{p}_i$ be an estimate of $p_i$. A binary prediction rule can be defined as follows: For a threshold $\theta\in\left[0,1\right]$, $Y_{i}$ is predicted as 
\begin{equation*}
\tilde{Y}_{i}(\theta)=\left\{
  \begin{array}{@{}ll@{}}
    0,&  \text{if}\ \hat{p}_{i}<\theta, \\
    1,&  \text{if}\ \hat{p}_{i}\geq\theta.
  \end{array}\right.
\end{equation*}
\textit{Sensitivity}, also known as the \textit{true positive rate} (TPR), is defined as the probability of correctly classifying a ``success''; and \textit{specificity}, 
which is equivalent to 
$1$ less the \textit{false positive rate} (FPR), is the probability of correctly classifying a ``failure''.
An ROC curve is a plot of \textit{sensitivity} versus $1-\textit{specificity}$ for all possible thresholds varying from 0 to 1. 
Let 
\begin{equation*}
    \boldsymbol{\hat{R}}(\theta)=(\hat{R}_{FP}(\theta),\hat{R}_{TP}(\theta))
\end{equation*}
be the FPR and TPR evaluated at threshold $\theta$ using the estimated probability of success. Then an empirical ROC curve can be defined as the function
\begin{equation*}
    \hat{f}=\{\boldsymbol{\hat{R}}(\theta):\theta\in[0,1]\}.
\end{equation*}

There are several summaries that can be obtained from an ROC curve, the most common of which is the area under the curve (AUC) that measures the ability of model predictions to discriminate between 
binary outcomes. AUC is equivalent to the c-statistic, which is the probability that a randomly selected ``success" observation has a higher predicted probability of being ``success" than a randomly selected ``failure" observation \citep{harrell2015regression}.
Furthermore, AUC is equivalent to the Mann-Whitney U-statistic 
\citep{bamber1975area}. 

\subsection{ROC analysis difficulties with paired comparisons}
\label{subsec:ROC-difficulties}
In most binary response settings, ROC analyses proceed in an unambiguous way because ``success'' and ``failure'' are clearly defined.
However, in paired comparison settings, 
the labeling of
``success'' and ``failure'' are relative to the team
that is considered the reference.
ROC analyses can no longer be performed in the usual way, 
and, depending on the labeling of success and failure, can 
result in different ROC analyses and inconsistent AUC summaries.

To illustrate the difficulty, consider
the Bradley-Terry model, as defined in (\ref{equa:bradley-terry}). For $k\in[n_{ij}]$,
\begin{equation*}
    Y_{ijk}\sim\text{Bernoulli}(p_{ij}),\text{ with}\ p_{ij}=\frac{1}{1+\exp(-\boldsymbol{x_{ij}}\boldsymbol{\mu})},
\end{equation*}
where $\boldsymbol{\mu}$ is an $m$-dimensional column vector of strength parameters. The pairing vector $\boldsymbol{x_{ij}}$ is an $m$-dimension row vector of the form
\begin{equation*}
    \boldsymbol{x_{ij}}=\begin{pmatrix}0,\dots,0,1,0,\dots,0,-1,0,\dots,0\end{pmatrix},
\end{equation*}
with $1$ at the $i^{th}$ entry, $-1$ at the $j^{th}$ entry, and $0$ everywhere else. 
Equivalently, the Bradley-Terry model can be written as,
\begin{equation*}
    1-Y_{ijk}\sim \text{Bernoulli}(1-p_{ij}),\text{ with}\ 1-p_{ij}=\frac{1}{1+\exp(\boldsymbol{x_{ij}}\boldsymbol{\mu})}.
\end{equation*}
The outcome of each competition between team $i$ and team $j$ can be recorded as either $Y_{ijk}$ or $1-Y_{ijk}$ depending on whether team $i$ is used as the reference. Hence for a data set containing $N$ games, there are $2^N$ different ways of recording the outcomes. Although inference for $\boldsymbol{\mu}$ via MLE remains the same no matter how the data are recorded, the TPRs and the FPRs can be different, leading to possibly different ROC curves. 
The ambiguity in the coding of game outcomes and the 
corresponding pairing vectors requires a refined
approach towards ROC analyses.

In the following sections,
we propose two novel approaches to perform ROC analysis for paired comparison data.
Each approach resolves the ambiguity in the paired comparison 
data recording.

\subsection{WL-ROC curves for paired comparison models}
\label{subsec:WL-ROC}
For the first approach, we define ``success" as winning a game and ``failure" as losing a game. Then each game could be both a ``success", if the winner is used as the reference, and a ``failure", if the loser is used as the reference. We will label each game in both ways to construct ROC curves that evaluate the ability of paired comparison models at discriminating between the winners and the losers. Hence we refer to the first approach as the Winner-Loser(WL)-ROC analysis, and the ROC curve constructed by this approach is called the Winner-Loser(WL)-ROC curve.

To construct the data set for analysis, first relabel each of the $N$ games using the winner as reference to obtain a data set with $N\times m$ design matrix 
\begin{equation*}
\boldsymbol{X^{w}}=\begin{pmatrix}\vdots\\\boldsymbol{x_{ijk}^{w}}\\\vdots\end{pmatrix},
\end{equation*} where $\boldsymbol{x_{ijk}^{w}}$ is the relabeled pairing vector for the $k^{th}$ game between team $i$ and team $j$ such that 
\begin{equation*}
 \boldsymbol{x_{ijk}^{w}}=\left\{
  \begin{array}{@{}ll@{}}
    \boldsymbol{x_{ij}}&  \text{if } Y_{ijk}=1,\\
    -\boldsymbol{x_{ij}}&  \text{if } Y_{ijk}=0. 
  \end{array}\right.
\end{equation*}
The corresponding vector of outcomes $\boldsymbol{Y^{w}}$ is an $N$-dimensional vector of ones indicating every game results in a win for the reference team. The next step is to relabel each game using the loser as reference and obtain a data set with $N\times m$ design matrix $\boldsymbol{X^{\ell}}=-\boldsymbol{X^{w}}$ and $N$-dimensional outcome vector $\boldsymbol{Y^{\ell}}$ being an $N$-dimensional vector of zeros. We therefore have two different representations of the original paired comparison data as a function of the winners and of the losers. By stacking $\boldsymbol{X^{w}}$ and $\boldsymbol{X^{\ell}}$ vertically into a $2N\times m$ design matrix and stacking $\boldsymbol{Y^{w}}$ and $\boldsymbol{Y^{\ell}}$ in the same way, we obtain a reconfigured data set consisting of $2N$ games, in which the first $N$ games are considered ``successes" and the rest $N$ games are considered ``failures". As we describe below, although we are doubling the size of the data set, the amount of information we use to perform WL-ROC analysis still remains the same so that the standard error is not inappropriately reduced. Recall that if $\boldsymbol{\hat{\mu}}$ are the estimates of the strength parameters, then the predicted probabilities of success for the ``successes" can be calculated as
\begin{equation*}
    \boldsymbol{\hat{p}^{w}}=F(\boldsymbol{X^w}\boldsymbol{\hat{\mu}}),
\end{equation*}
and the predicted probabilities of success for the ``failures" can be calculated as
\begin{equation*}
    \boldsymbol{\hat{p}^{\ell}}=F(\boldsymbol{X^{\ell}}\boldsymbol{\hat{\mu}}).
\end{equation*}
Since ``success" and ``failure" are clearly defined, the WL-ROC curve for paired comparison models on this 
reconfigured data set can be constructed in the same way 
as for ordinary binary response models, as follows.

For $i,j\in[m]$ and $i<j$, let $\hat{p}_{ij}$ be the estimate of $p_{ij}$, and we can arrange all $\hat{p}_{ij}$ and $1-\hat{p}_{ij}$ in non-decreasing order as $\boldsymbol{\hat{p}}=\big(\hat{p}_{(r)}\big)_{r\in[2R]}$. We can then arrange all $w_{ij}$ and $n_{ij}-w_{ij}$ according to $\boldsymbol{\hat{p}}$ as $\boldsymbol{w_{\hat{p}}}=\big(w_{\hat{p}_{(r)}}\big)_{r\in[2R]}$. Then the estimated FPR and TPR at threshold $\theta$ can be written as
\begin{align*}
\hat{R}_{FP}^{w\ell}(\theta) &= \frac{\sum_{r=1}^{2R}w_{\hat{p}_{(2R+1-r)}}\mathbb{1}_{\{\hat{p}_{(r)}>\theta\}}}{N}, \\
\hat{R}_{TP}^{w\ell}(\theta) &= \frac{\sum_{r=1}^{2R}w_{\hat{p}_{(r)}}\mathbb{1}_{\{\hat{p}_{(r)}>\theta\}}}{N}. 
\end{align*}
Therefore, the WL-ROC curve can be defined as a piecewise linear function $\widehat{f}^{w\ell}$ with the knots $\boldsymbol{\hat{R}^{w\ell}}$ being all possible pairs of FPR and TPR such that
\begin{equation}
\label{equa:estimatedRwl}
\boldsymbol{\hat{R}^{w\ell}} =\Big\{(0,0)\Big\}\cup\Big\{\big(\frac{\sum_{r=1}^{2R-k+1}w_{\hat{p}_{(r)}}}{N},\frac{\sum_{r=k}^{2R}w_{\hat{p}_{(r)}}}{N}\big)\Big\}_{k\in[2R]}.
\end{equation}

Interpretations of the area under the WL-ROC curve (AUWLC) are similar to those of the AUC for ordinary binary response models, as discussed in section~\ref{subsec:ROCbin}. Define WL-c-statistic as the probability that a randomly selected winner has a higher predicted probability of winning than a randomly selected loser. Considering all possible pairs of $\hat{p}^{w}_{a}\in\boldsymbol{\hat{p}^{w}}$ and $\hat{p}^{\ell}_{b}\in\boldsymbol{\hat{p}^{\ell}}$, the WL-c-statistic, denoted as $\hat{c}^{w\ell}$, is calculated as the proportion of pairs such that $\hat{p}^{w}_{a}$ is greater than $\hat{p}^{\ell}_{b}$; that is, 
\begin{equation}
\label{equa:estimatedCwl}
    \hat{c}^{w\ell}=\frac{\sum_{a,b}\big(\mathbb{1}_{\{\hat{p}^{w}_a>\hat{p}^{\ell}_b\}}+\frac{1}{2}\mathbb{1}_{\{\hat{p}^{w}_a=\hat{p}^{\ell}_b\}}\big)}{N^2}.
\end{equation}
Notice that the WL-c-statistic is calculated in the same way as the Mann-Whitney U-statistic. Hence AUWLC is equivalent to the WL-c-statistic and the Mann-Whitney U-statistic between $\boldsymbol{\hat{p}^{w}}$ and $\boldsymbol{\hat{p}^{\ell}}$.

\subsection{SW-ROC curves for paired comparison models} 
\label{subsec:SW-ROC}
The second approach we propose involves configuring the data set using estimates of the strength parameters. Of the two teams involved in each game, the team with higher strength estimate is considered the stronger team, and the other team is the weaker team. In Section~\ref{subsec:sw-ties}, we address the situation when pairs of teams have equal strength estimates.  We define ``success" as a game such that the stronger team defeats the weaker team and define ``failure" as a game such that the weaker team wins. Henceforth, we will call the stronger team that wins as a strong winner and the weaker team that wins as a weak winner. Our second approach is to construct ROC curves that evaluate the ability of paired comparison models in discriminating stronger winners and weak winners, and thus we call this approach the Strong-Weak(SW)-ROC analysis and the corresponding ROC curves the Strong-Weak(SW)-ROC curves. 

To perform SW-ROC analysis, relabel each game using the stronger team as the reference, and then the relabeled $N\times m$ design matrix is \begin{equation*}
\boldsymbol{\hat{X}^{sw}}=\begin{pmatrix}\vdots\\\boldsymbol{\hat{x}_{ijk}^{sw}}\\\vdots\end{pmatrix},
\end{equation*} where $\boldsymbol{\hat{x}_{ijk}^{sw}}$ is the pairing vector for the $k^{th}$ game between team $i$ and team $j$ labeled according to the stronger team such that 
\begin{equation*}
 \boldsymbol{\hat{x}_{ijk}^{sw}}=\left\{
  \begin{array}{@{}ll@{}}
    \boldsymbol{x_{ij}}&  \text{if } \hat{\mu}_i>\hat{\mu}_j,\\
    -\boldsymbol{x_{ij}}&  \text{if } \hat{\mu}_i<\hat{\mu}_j. 
  \end{array}\right.
\end{equation*}
The outcome vector $\boldsymbol{\hat{Y}^{sw}}$ is relabeled accordingly such that 
\begin{equation*}
\hat{Y}_{ijk}^{sw}=\left\{
  \begin{array}{@{}ll@{}}
    1&  \text{if }Y_{ijk}=1 \text{ and } \hat{\mu}_i>\hat{\mu}_j\text{, or if }Y_{ijk}=0 \text{ and } \hat{\mu}_i<\hat{\mu}_j,\\
    0&  \text{otherwise}.
  \end{array}\right.
\end{equation*}
The reconfigured data set consists of $\hat{W}=\sum_{i,j,k}Y_{ijk}^{sw}$ games with strong winners as ``successes" and $N-\hat{W}$ games with weak winners as ``failures". The predicted probabilities of winning for the stronger teams involved in the $N$ games are calculated as 
\begin{equation*}
\boldsymbol{\hat{p}}^{sw}=F(\boldsymbol{X^{sw}}\boldsymbol{\hat{\mu}}),
\end{equation*}
which consists of the predicted probabilities of winning for the stronger winners $\boldsymbol{\hat{p}^{st}}$ and the predicted probabilities of winning for the weak winners $\boldsymbol{\hat{p}^{wk}}$. Considering $\boldsymbol{\hat{p}^{st}}$ and $\boldsymbol{\hat{p}^{wk}}$ as the probability of success for the ``successes" and the ``failures", the SW-ROC curve can be constructed in the usual way.

The SW-ROC curve can also be defined as a piecewise linear function. For $i,j\in[m]$ and $i<j$, let $\hat{q}_{ij}=\text{max}\{\hat{p}_{ij},1-\hat{p}_{ij}\}$ be the estimated probability that the stronger team between team $i$ and team $j$ wins, and arrange all $\hat{q}_{ij}$ in non-decreasing order as $\boldsymbol{\hat{q}}=\big(\hat{q}_{(r)}\big)_{r\in[R]}$. Define $\boldsymbol{n_{\hat{q}}}=\big(n_{\hat{q}_{(r)}}\big)_{r\in[R]}$ and $\boldsymbol{w_{\hat{q}}}=\big(w_{\hat{q}_{(r)}}\big)_{r\in[R]}$ such that $n_{\hat{q}_{(r)}}$ is the number of comparisons between the pair of teams corresponding to $\hat{q}_{(r)}$, and $w_{\hat{q}_{(r)}}$ is the number of wins by the stronger team. As defined in section~\ref{subsec:WL-ROC}, $\boldsymbol{w_{\hat{p}}}=\big(w_{\hat{p}_{(r)}}\big)_{r\in[2R]}$, so $\boldsymbol{n_{\hat{q}}}=\big(w_{\hat{p}_{(r)}}\big)_{r\in\{R+1,\dots,2R\}}$. The total number of wins by the stronger teams $\hat{W}$ can be calculated as $\hat{W}=\sum_{r=1}^{R}w_{\hat{q}_{(r)}}$. For a threshold $\theta$, the estimated FPR and TPR are 
\begin{align*}
\hat{R}_{FP}^{sw}(\theta) &= \frac{\sum_{r=1}^{R}(n_{\hat{q}_{(r)}}-w_{\hat{q}_{(r)}})\mathbb{1}_{\{\hat{q}_{(r)}>\theta\}}}{N-\hat{W}},\\
\hat{R}_{TP}^{sw}(\theta) &= \frac{\sum_{r=1}^{R}w_{\hat{q}_{(r)}}\mathbb{1}_{\{\hat{q}_{(r)}>\theta\}}}{\hat{W}}.
\end{align*}
Hence we can write the SW-ROC curve as piecewise linear function $\widehat{f}^{sw}$ with knots 
\begin{equation}
\label{equa:estimatedRsw}
\boldsymbol{\hat{R}^{sw}}=\Big\{(0,0)\Big\}\cup\Big\{\big(\frac{\sum_{r=k}^{R}n_{\hat{q}_{(r)}}-w_{\hat{q}_{(r)}}}{N-\hat{W}},\frac{\sum_{r=k}^{R}w_{\hat{q}_{(r)}}}{\hat{W}}\big)\Big\}_{k\in[R]}.
\end{equation}

To interpret the area under an SW-ROC curve (AUSWC), we define the SW-c-statistic as the probability that a randomly selected stronger winner has a higher predicted probability of winning than a randomly selected weak winner. Hence the SW-c-statistic is calculated as
\begin{equation}
\label{equa:estimatedCsw}
    \hat{c}^{sw}=\frac{\sum_{a,b}\big(\mathbb{1}_{\{\hat{p}^{st}_a>\hat{p}^{wk}_b\}}+\frac{1}{2}\mathbb{1}_{\{\hat{p}^{st}_a=\hat{p}^{wk}_b\}}\big)}{\hat{W}(N-\hat{W})}.
\end{equation}
Notice that the SW-c-statistic is defined in a similar way as the WL-c-statistic with changes to fit the definition of ``success" and ``failure" in the SW-ROC analysis setting. It then follows that AUSWC is equivalent to the SW-c-statistic and the Mann-Whitney U-statistic between $\boldsymbol{\hat{p}^{st}}$ and $\boldsymbol{\hat{p}^{wk}}$. 

\subsubsection{SW-ROC curves with ties in strength estimates}
\label{subsec:sw-ties}
Now we consider the situation where some of the strength parameter estimates are equal and describe how SW-ROC curves can be constructed under this situation. Suppose team $i$ and team $j$ have equal strength estimates, with $\hat{\mu}_{i}=\hat{\mu}_{j}$. In our analysis, we arbitrarily pick one of the two teams as the reference (stronger) team and assign that team to have $n_{ij}/2$ wins.

More generally, allowing for more than two teams with equal estimated strength parameters, assume there are $m_0$ distinct elements of $\boldsymbol{\hat{\mu}}$ and that we can split the $m$ teams into $m_0$ groups $S_1,S_2,\dots,S_{m_0}$ such that the teams with equal strength estimates are in the same group. Let
\begin{equation*}
    R_{0}=\sum_{k=1}^{m_0}\frac{|S_k|(|S_k|-1)}{2}
\end{equation*}
be the number of pairs with equal strength estimates, and let
\begin{equation*}
    N_0=\sum_{k:|S_k|\geq2}\ \sum_{i,j\in S_k}\frac{n_{ij}}{2}
\end{equation*}
be the reassigned number of wins by the stronger teams of the pairs with equal strength estimates, then 
\begin{equation*}
    \hat{W}=N_0 +\sum_{r=R_0+1}^{R}w_{\hat{q}_{(r)}}.
\end{equation*} 
The SW-ROC curve is still piecewise linear with knots
\begin{equation*}
\Big\{(0,0),(1,1)\Big\}\cup\Big\{\big(\frac{\sum_{r=k}^{R}n_{\hat{q}_{(r)}}-w_{\hat{q}_{(r)}}}{N-\hat{W}},\frac{\sum_{r=k}^{R}w_{\hat{q}_{(r)}}}{\hat{W}}\big)\Big\}_{k\in\{R_0+1,\dots,R\}}.
\end{equation*}

\subsection{Asymptotic properties of WL-ROC curves and SW-ROC curves}
\label{subsec:Asymptotic}
In Section~\ref{subsec:WL-ROC} and Section~\ref{subsec:SW-ROC}, we defined the WL-ROC curve $\widehat{f}^{w\ell}$ and the SW-ROC curve $\widehat{f}^{sw}$ respectively. For each estimated ROC curve, there also exists a true ROC curve if we were to construct it using the true strength parameters $\boldsymbol{\mu}$ which are unknown. We can describe the true ROC curves in the same way as the estimated ROC curves by defining $\boldsymbol{p}$, $\boldsymbol{w_p}$, $\boldsymbol{q}$, $\boldsymbol{n_q}$, $\boldsymbol{w_q}$, and $W$ accordingly using the true strength parameters. Then the true WL-ROC curve can be written as a piecewise linear function $f^{w\ell}$ with knots
\begin{equation}
\label{equa:trueRwl}
\boldsymbol{R^{w\ell}} =\Big\{(0,0)\Big\}\cup\Big\{\big(\frac{\sum_{r=1}^{2R-k+1}w_{p_{(r)}}}{N},\frac{\sum_{r=k}^{2R}w_{p_{(r)}}}{N}\big)\Big\}_{k\in[2R]}.
\end{equation}
Similarly, the true SW-ROC curve can be considered as a piecewise linear function $f^{sw}$ with knots 
\begin{equation}
\label{equa:trueRsw}
\boldsymbol{R^{sw}}=\Big\{(0,0)\Big\}\cup\Big\{\big(\frac{\sum_{r=k}^{R}n_{q_{(r)}}-w_{q_{(r)}}}{N-W},\frac{\sum_{r=k}^{R}w_{q_{(r)}}}{W}\big)\Big\}_{k\in[R]}.
\end{equation}

Both the estimated curves and true curves are constructed on the given data set, which corresponds to a specific design matrix. Suppose the teams continue to compete for infinitely many games according to a design $\boldsymbol{\mathcal{D}}$, then there exist a limiting true WL-ROC curve and a limiting true SW-ROC curve which are functions of the design $\boldsymbol{\mathcal{D}}$ and the true strength parameters $\boldsymbol{\mu}$. For $r\in[R]$, let $d_r=n_{q_{(r)}}/N$ be the ratio of the number of comparisons between the two teams corresponding to $q_{(r)}$ to the total number of comparisons, then $\boldsymbol{\mathcal{D}}$ can be characterized by the ratios as $\boldsymbol{\mathcal{D}}=(d_r)_{r\in[R]}$. Notice that although the ratios are defined with respect to $\boldsymbol{q}$, they can also be extended for $\boldsymbol{p}$. Let $\boldsymbol{n_{p}}$ be the number of comparisons corresponding to $\boldsymbol{p}$. By the fact that
\begin{align*}
    \boldsymbol{n_{p}}=(n_{p_{(r)}})_{r\in[2R]}&=(n_{q_{(R)}},\dots,n_{q_{(1)}},n_{q_{(1)}},\dots,n_{q_{(R)}}) \\
    &=(n_{q_{(|r-R-1/2|+1/2)}})_{r\in[2R]},
\end{align*}
the design ratios corresponding to $\boldsymbol{n_p}$ can be written as $(d_{|r-R-1/2|+1/2})_{r\in[2R]}$. Since the ratios are fixed by design, the limiting true WL-ROC curve can be defined as a piecewise linear function $\widetilde{f}^{w\ell}_\mathcal{D}$ with knots
\begin{equation}
\label{equa:limittrueRwl}
\boldsymbol{\tilde{R}^{w\ell}} =\Big\{(0,0)\Big\}\cup\Big\{\big(\sum_{r=1}^{2R-k+1}p_{(r)}d_{|r-R-\frac{1}{2}|+\frac{1}{2}},\sum_{r=k}^{2R}p_{(r)}d_{|r-R-\frac{1}{2}|+\frac{1}{2}}\big)\Big\}_{k\in[2R]}.
\end{equation}
The limiting true SW-ROC curve is a piecewise linear function $\widetilde{f}^{sw}_\mathcal{D}$ with knots 
\begin{equation}
\label{equa:limittrueRsw}
\boldsymbol{\tilde{R}^{sw}}=\Big\{(0,0)\Big\}\cup\Big\{\big(\frac{\sum_{r=k}^{R}(1-q_{(r)})d_{r}}{\sum_{r=1}^{R}(1-q_{(r)})d_{r}},\frac{\sum_{r=k}^{R}q_{(r)}d_{r}}{\sum_{r=1}^{R}q_{(r)}d_{r}}\big)\Big\}_{k\in[R]}.
\end{equation}

Before stating the asymptotic behavior of WL-ROC curve and SW-ROC curve, we first introduce the ROC metric that we use to measure the distance between two ROC curves. As proposed in \cite{alsing2002convergence}, for two ROC curves $f$ and $g$ such that
\begin{align*}
    f &= \Big\{(R_{FP}^{f}(\theta),R_{TP}^{f}(\theta)):\theta\in[0,1]\Big\} = \Big\{\boldsymbol{R^{f}}(\theta):\theta\in[0,1]\Big\}, \\
    g &= \Big\{(R_{FP}^{g}(\theta),R_{TP}^{g}(\theta)):\theta\in[0,1]\Big\} = \Big\{\boldsymbol{R^{g}}(\theta):\theta\in[0,1]\Big\},
\end{align*}
the ROC metric $d_{\rho,z}$ is defined as 
\begin{equation}
\label{equa:ROCmetric}
    d_{\rho,z}(f,g)=\bigg(\int_{0}^{1}\rho\big(\boldsymbol{R^{f}}(\theta),\boldsymbol{R^{g}}(\theta)\big)^{z}d\theta\bigg)^{1/z},
\end{equation}
where $1\leq z<\infty$ and $\rho$ is a metric on $\mathbb{R}^2$. The main results of the convergence of WL-ROC curves and SW-ROC curves are established through the ROC metric defined in (\ref{equa:ROCmetric}) and formulated in Theorem~\ref{theorem:convergence}.
\begin{theorem}
\label{theorem:convergence}
Let $\{\widehat{f}^{w\ell}_{\boldsymbol{n}}\}$ and $\{f^{w\ell}_{\boldsymbol{n}}\}$ be a sequence of estimated WL-ROC curves and a sequence of corresponding true WL-ROC curves indexed by the numbers of pairwise comparisons $\boldsymbol{n}=\{n_{q_{(r)}}\}_{r\in[R]}$. Let $\{\widehat{f}^{sw}_{\boldsymbol{n}}\}$ and $\{f^{sw}_{\boldsymbol{n}}\}$ be a sequence of estimated SW-ROC curves and a sequence of true SW-ROC curves indexed by $\boldsymbol{n}$. For $\epsilon>0$, there exists $N^{*}$ such that for all nonzero $n_{q_{(r)}}>N^{*}$,
\begin{align*}
    P\Big(d_{\rho,z}\big(\widehat{f}^{w\ell}_{\boldsymbol{n}},f^{w\ell}_{\boldsymbol{n}}\big)\geq\epsilon\Big)<\epsilon,\ \ 
    P\Big(d_{\rho,z}\big(\widehat{f}^{sw}_{\boldsymbol{n}},f^{sw}_{\boldsymbol{n}}\big)\geq\epsilon\Big)<\epsilon,
\end{align*}
and for all nonzero $n_{q_{(r)}}>N^{*}$ with the design $\boldsymbol{\mathcal{D}}$,
\begin{align*}
    P\Big(d_{\rho,z}\big(\widehat{f}^{w\ell}_{\boldsymbol{n}},\widetilde{f}^{w\ell}_{\boldsymbol{\mathcal{D}}}\big)\geq\epsilon\Big)<\epsilon,\ \ 
    P\Big(d_{\rho,z}\big(\widehat{f}^{sw}_{\boldsymbol{n}},\widetilde{f}^{sw}_{\boldsymbol{\mathcal{D}}}\big)\geq\epsilon\Big)<\epsilon.
\end{align*}
\end{theorem}
Let us first consider the convergence behaviors of WL-ROC curves. As the teams continue to compete for more and more games, the estimated WL-ROC curve will converge to the true WL-ROC curve in probability under the ROC metric defined in (\ref{equa:ROCmetric}). If we further assume that the number of games goes to infinity with fixed ratios by design $\boldsymbol{\mathcal{D}}$, then the estimated WL-ROC curve will also converge to the limiting true WL-ROC curve in probability under the ROC metric. Similarly, the SW-ROC curve also exhibits such convergence behaviors. The proof of Theorem~\ref{theorem:convergence} is included in the Appendix~\ref{app:convergence}. 

\section{WL-c-statistic and SW-c-statistic}
\label{sec:comparison}
The WL-c-statistic $\hat{c}^{w\ell}$ defined in section~\ref{subsec:WL-ROC} and the SW-c-statistic $\hat{c}^{sw}$ defined in section~\ref{subsec:SW-ROC} both summarize the discrimination ability of paired comparison models. The difference is that $\hat{c}^{w\ell}$ evaluates the ability to discriminate between winners and losers, while $\hat{c}^{sw}$ evaluates the ability to discriminate between strong winners and weak winners. To establish relationships between $\hat{c}^{w\ell}$ and $\hat{c}^{sw}$, we will first redefine the WL-ROC curve using the notations for the SW-ROC curve. In fact, 
$\boldsymbol{\hat{p}}$ and $\boldsymbol{w_{\hat{p}}}$ can be written in terms of $\boldsymbol{\hat{q}}$, $\boldsymbol{w_{\hat{q}}}$, and $\boldsymbol{n_{\hat{q}}}$ such that
\begin{align*}
    \boldsymbol{\hat{p}} &= \Big(1-\hat{q}_{(R)}, \dots, 1-\hat{q}_{(1)}, \hat{q}_{(1)}, \dots, \hat{q}_{(R)}\Big), \\
    \boldsymbol{w_{\hat{p}}} &= \Big(n_{\hat{q}_{(R)}}-w_{\hat{q}_{(R)}}, \dots, n_{\hat{q}_{(1)}}-w_{\hat{q}_{(1)}}, w_{\hat{q}_{(1)}}, \dots, w_{\hat{q}_{(R)}}\Big).
\end{align*}
Therefore, the knots $\boldsymbol{R^{w\ell}}$ of the WL-ROC curve defined in (\ref{equa:estimatedRwl}) can be rewritten in terms of $\boldsymbol{\hat{q}}$, $\boldsymbol{w_{\hat{q}}}$, and $\boldsymbol{n_{\hat{q}}}$. Using these notations for both the WL-ROC curve and the SW-ROC curve, we will compare $\hat{c}^{w\ell}$ with $\hat{c}^{sw}$ and discuss their standard errors in the remainder of this section.

\subsection{Linear relationship between WL-c-statistic and SW-c-statistic}
\label{subsec:linear}
We now demonstrate that a linear relationship exists between $\hat{c}^{w\ell}$ and $\hat{c}^{sw}$. Recall that $N$ denotes the total number of games and $\hat{W}$ denotes the number of games that the estimated stronger team wins, then
\begin{equation}
\label{equa:linear}
    N^2\hat{c}^{wl}=2\hat{W}(N-\hat{W})\hat{c}^{sw}+(\hat{W})^2.
\end{equation}
This relationship is established using the fact that
\begin{equation}
\label{equa:linear-estimatedCwl}
    \hat{c}^{w\ell} = \frac{2\sum_{r=1}^{R}\hat{A}_rw_{\hat{q}_{(r)}}}{N^2},
\end{equation}
and
\begin{equation}
\label{equa:linear-estimatedCsw}
    \hat{c}^{sw} = \frac{\sum_{r=1}^{R}\hat{A}_rw_{\hat{q}_{(r)}}-\frac{1}{2}(\hat{W})^2}{\hat{W}(N-\hat{W})},
\end{equation}
where $\hat{A}_r=\sum_{s=1}^{r-1}n_{\hat{q}_{(s)}}+\frac{1}{2}n_{\hat{q}_{(r)}}$. The proof of (\ref{equa:linear}) is presented in Appendix~\ref{app:linear}.

\subsection{Comparison between WL-c-statistic and SW-c-statistic}
Define $\tilde{c}^{w\ell}$ and $\tilde{c}^{sw}$ as the limiting WL-c-statistic and limiting SW-c-statistic corresponding to the limiting true WL-ROC curve and limiting true SW-ROC curve, as defined in (\ref{equa:limittrueRwl}) and (\ref{equa:limittrueRsw}), respectively. Assume that each pair of teams compete for exactly $n$ times so $n_{ij}=n$ for $i,j\in[m]$ and $i<j$, then it follows from (\ref{equa:linear-estimatedCwl}) and (\ref{equa:linear-estimatedCsw}) that 
\begin{align}
    \label{equa:limitinngCwl}
    \tilde{c}^{w\ell}
    &=\frac{2\sum_{r=1}^{R}(r-\frac{1}{2})q_{(r)}}{R^2},\\
    \label{equa:limitinngCsw}
    \tilde{c}^{sw}
    &= \frac{\sum_{r=1}^{R}(r-\frac{1}{2})q_{(r)}-\frac{1}{2}\big(\sum_{r=1}^{R}q_{(r)}\big)^2}{\big(\sum_{r=1}^{R}q_{(r)}\big)\big(R-\sum_{r=1}^{R}q_{(r)}\big)}.
\end{align}
The limiting WL-c-statistic is always greater than or equal to the limiting SW-c-statistic; that is, 
\begin{equation}
\label{equa:c-compare}
    \tilde{c}^{w\ell}\geq \tilde{c}^{sw},
\end{equation} where the equality holds only when all teams have equal true strength parameters. Hence the paired comparison models perform asymptotically better at discriminating between winners and losers than discriminating between strong winners and weak winners. The proof of (\ref{equa:c-compare}) is presented in Appendix~\ref{app:c-compare}.

\subsection{Standard errors of WL-c-statistic and SW-c-statistic}
\begin{figure}[ht]
    \centering
    \includegraphics[scale=0.6]{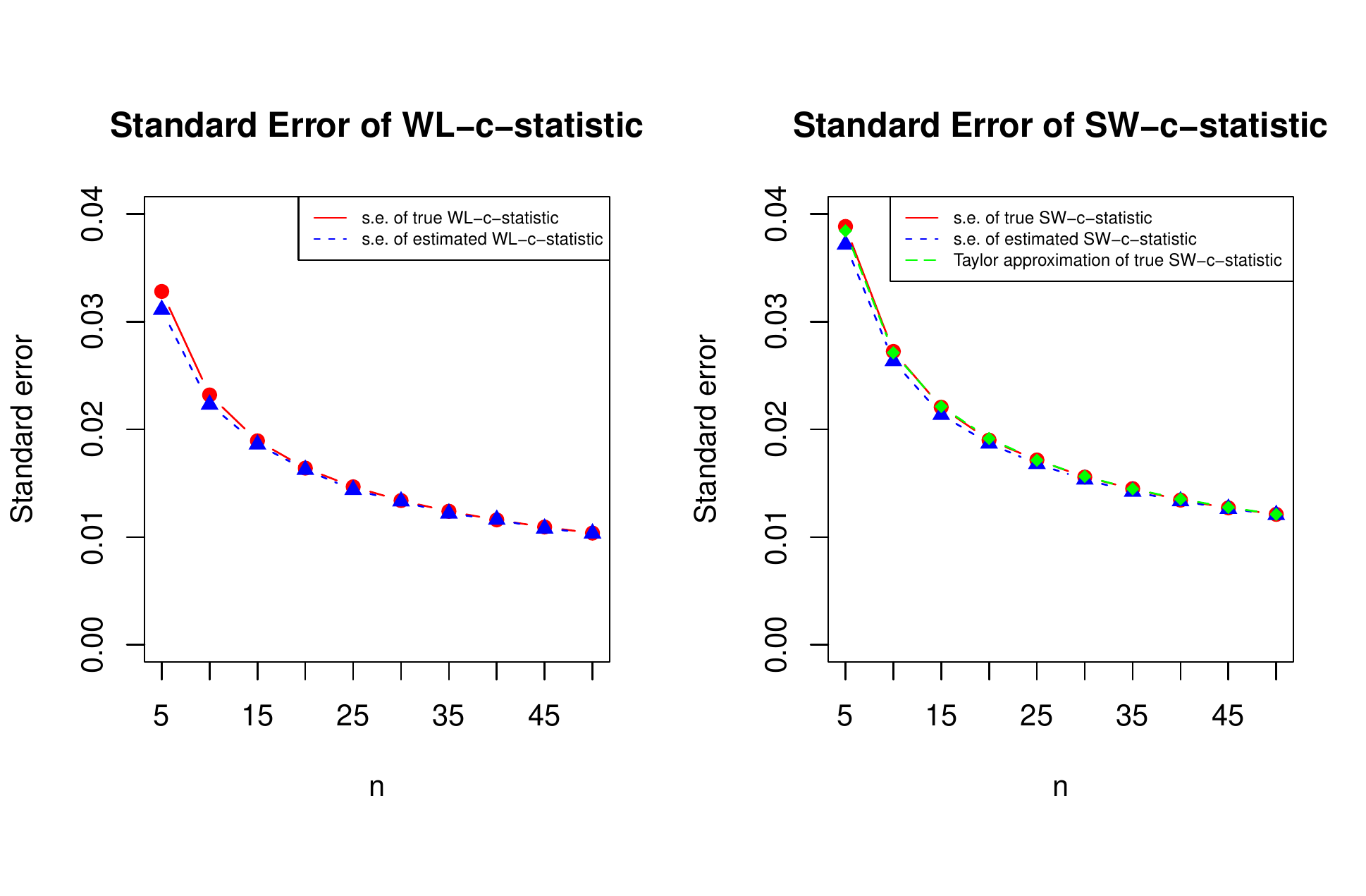}
    \caption{The standard errors of WL-c-statistic and SW-c-statistic versus $n$, the number of games within each pair of teams, based on simulated paired comparison data with 10 teams.}
    \label{Figure:SE}
\end{figure}
Rather than finding exact closed-form solutions of the standard errors of $\hat{c}^{w\ell}$ and $\hat{c}^{sw}$ which are complicated by the estimated winning probabilities $\boldsymbol{\hat{q}}$ as part of the expressions, we can obtain closed-form solutions or approximations of the standard errors of the true c-statistics. First, let us define the true WL-c-statistic $c^{w\ell}$ and true SW-c-statistic $c^{sw}$ as the c-statistics corresponding to the true WL-ROC curve and true SW-ROC curve defined in (\ref{equa:trueRwl}) and (\ref{equa:trueRsw}). By replacing the estimates in (\ref{equa:linear-estimatedCwl}) and (\ref{equa:linear-estimatedCsw}) with the true parameters, we get
\begin{align}
    c^{w\ell}&=\frac{2\sum_{r=1}^{R}A_rw_{q_{(r)}}}{N^2}, \\
    c^{sw}&=\frac{\sum_{r=1}^{R}A_rw_{q_{(r)}}-\frac{1}{2}(W)^2}{W(N-W)},
\end{align}
where $A_r=\sum_{s=1}^{r-1}n_{q_{(s)}}+\frac{1}{2}n_{q_{(r)}}$. For fixed $\boldsymbol{q}$ and $\boldsymbol{n_{q}}$, the standard error of $c^{w\ell}$ can be calculated as
\begin{align}
\label{equa:se-Cwl}
     \text{s.e.}(c^{w\ell})=\frac{2}{N^2}\sqrt{\sum_{r=1}^{R}A_r^2n_{q_{(r)}}q_{(r)}(1-q_{(r)})}. 
\end{align}
Deriving the standard error of $c^{sw}$ acknowledges that both the numerator and the denominator of $c^{sw}$ are random but we can approximate the standard error of $c^{sw}$ using Taylor expansions. Given two random variables $X$ and $Y$ with means ($u_X$, $u_Y$), variances $(\sigma_X^2,\sigma_Y^2)$, and covariance $\sigma_{XY}$, the variance of $X/Y$ can be approximated using the first order Taylor expansion around $(\mu_X,\mu_Y)$:
\begin{equation*}
    \text{Var}\Big(\frac{X}{Y}\Big)\approx\frac{u_X^2}{u_Y^2}\Big(\frac{\sigma_X^2}{\mu_X^2}-\frac{2\sigma_{XY}}{\mu_X\mu_Y}+\frac{\sigma_Y^2}{\mu_Y^2}\Big).
\end{equation*}
To approximate the standard error of $c^{sw}$, we let $X =\sum_{r=1}^{R}A_rw_{q_{(r)}}-\frac{1}{2}(W)^2$ and $Y = W(N-W)$. Then we just need to find $\mu_X$, $\mu_Y$, $\sigma_X^2$, $\sigma_Y^2$, and $\sigma_{XY}$.
Let 
$B_1=E[W]$, $B_2=E[W^2]$, 
$B_3=E[W^3]$, and $B_4=E[W^4]$. Also, let $C_1=E\big[\sum_{r=1}^{R}A_rw_{q_{(r)}}\big]$ and $C_2=E\big[(\sum_{r=1}^{R}A_rw_{q_{(r)}})^2\big]$. Furthermore, let $D_1=E\big[W\sum_{r=1}^{R}A_rw_{q_{(r)}}\big]$ and $D_2=E\big[W^2\sum_{r=1}^{R}A_rw_{q_{(r)}}\big]$. Then the variance of $c^{sw}$ can be approximated as
\begin{align}
\label{equa:se-Csw}
    \text{s.e.}(c^{sw})\approx\frac{u_X}{u_Y}\sqrt{\frac{\sigma_X^2}{\mu_X^2}-\frac{2\sigma_{XY}}{\mu_X\mu_Y}+\frac{\sigma_Y^2}{\mu_Y^2}},
\end{align}
where 
\begin{align*}
    &\mu_X=C_1-\frac{1}{2}B_2,\ \ 
    \sigma_X^2=C_2-D_2+\frac{1}{4}B_4-\mu_X^2,\\
    &\mu_Y=NB_1-B_2,\ \ 
    \sigma_Y^2=N^2B_2-2NB_3+B_4-\mu_Y^2,\\
    &\sigma_{XY}=ND_1-D_2-\frac{1}{2}NB_3+\frac{1}{2}B_4-\mu_X\mu_Y.
\end{align*}

In Figure~\ref{Figure:SE}, we present the results of our paired comparison simulations involving $10$ teams. We assume that each pair of teams compete for $n$ times and increase $n$ from 5 to 50 in increments of 5. The standard errors of $c^{w\ell}$, $\hat{c}^{w\ell}$, $c^{sw}$, $\hat{c}^{sw}$, and the Taylor approximation of the standard error of $c^{sw}$ are computed for each $n$. The standard errors of $c^{wl}$ and the Taylor approximation of the standard error of $c^{sw}$ are calculated through (\ref{equa:se-Cwl}) and (\ref{equa:se-Csw}) respectively. The standard errors of $\hat{c}^{w\ell}$, $\hat{c}^{sw}$, and $c^{sw}$ are calculated based on simulation results. As shown in Figure~\ref{Figure:SE}, the standard errors all shrink to $0$ at the rate of $n^{-1/2}$.

\section{Applications}
\label{sec:applications}
\begin{figure}
    \centering
    \includegraphics[scale=0.7]{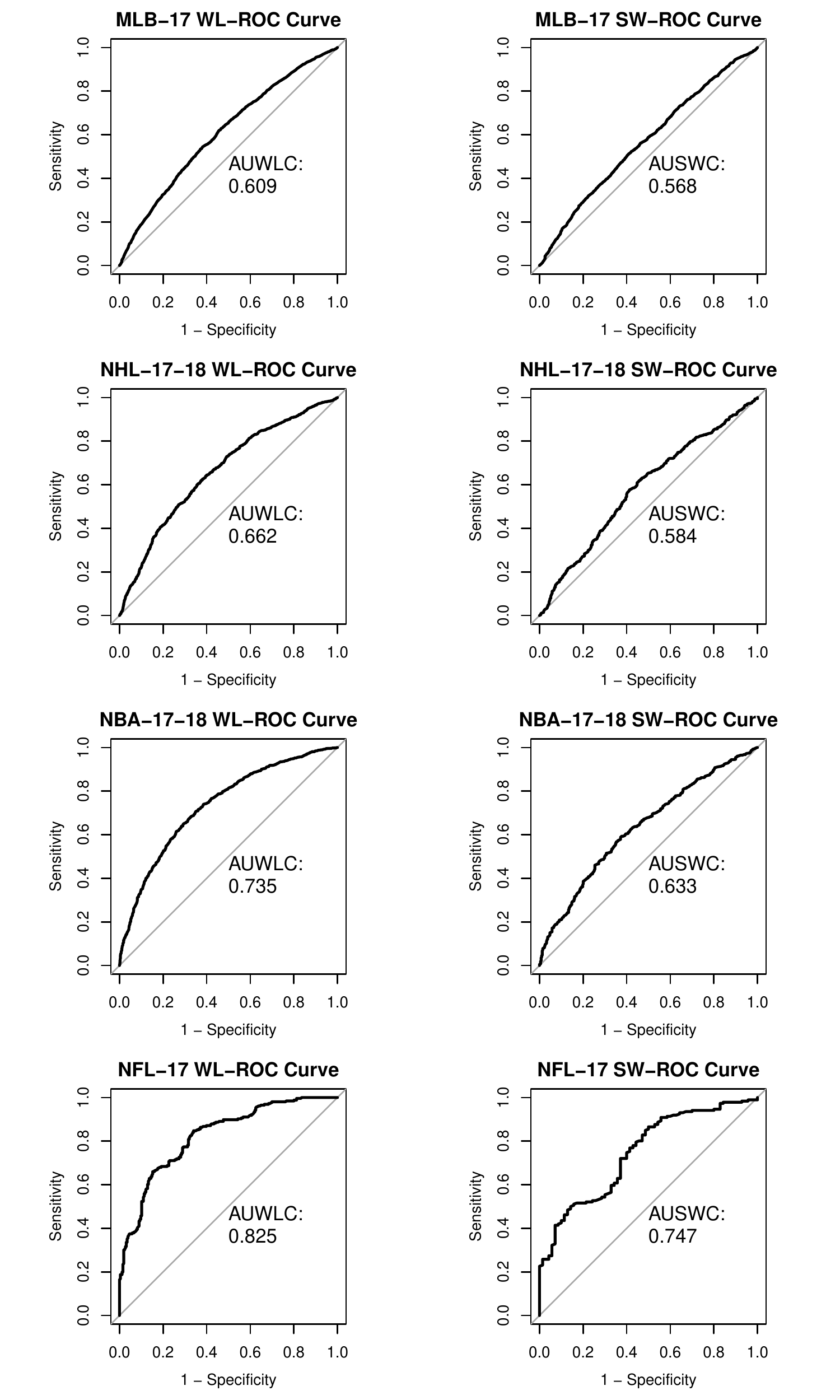}
    \caption{The WL-ROC curves and SW-ROC curves for Bradley-Terry model on 2017(-18) regular-season MLB, NHL, NBA, and NFL game outcomes.}
    \label{fig:4sports}
\end{figure}

In this section, we apply WL-ROC analysis and SW-ROC analysis to historical games played in MLB, NBA, NHL, and NFL (data source: \url{http://www.shrpsports.com}). The MLB data are from the 2017 regular-season involving 30 teams and 2430 games. The NHL data are from the 2017-18 regular-season involving 31 games and 1271 game. For the NFL games, we use the outcomes from the 2017 season consisting of 256 games among 32 teams. Finally, the NBA data are from the 2017-18 regular-season involving 30 teams and 1230 games. We fit the Bradley-Terry model for each league and construct the corresponding WL-ROC curve and SW-ROC curve, as shown in Figure~\ref{fig:4sports}.

First, let us compare the WL-ROC curve with the SW-ROC curve for each league. For each of the four leagues, the AUWLC is greater than the AUSWC, which implies that the Bradley-Terry model performs better at discriminating between winner and losers than discriminating between strong winners and weak winners on the given game outcomes. Comparing the ROC curves across the four leagues, we can see that the AUWLC and AUSWC for MLB and NHL are lower than those for NFL and NBA. It is noticeable that the ROC curves for MLB are very close to the diagonal line, meaning that the Bradley-Terry model performs just slightly better than random guessing at predicting baseball game outcomes. This could be related to a high level of parity, or unpredictability, among MLB games. In fact, for the 2017 season MLB, the team predicted to be the strongest (Cleveland Indians) only won $63\%$ of the games, while for the 2017-18 season NBA, the team predicted to be the strongest (Houston Rockets) won $80\%$ of their games.

To further explore the level of parity across different leagues, we calculate the relative standard deviation (RSD) for each league as a comparison metric \citep{fort2017competitive}. RSD is calculated as the ratio of actual standard deviation (ASD) to idealized standard deviation (ISD), where ASD is the standard deviation of actual end-of-season outcomes such as win percentages or points, and ISD is the standard deviation of end-of-season outcomes assuming all teams are equally likely to win. For a league in which $G$ games are scheduled for each team, the ISD of end-of-season win percentages is $0.5/\sqrt{G}$ \citep{fort1995cross}. The results based on end-of-season win percentages are shown in Table~\ref{table:1}. 

Since the level of parity and the RSD are inversely related, the results in Table~\ref{table:1} imply that NFL has the highest level of parity while NBA has the lowest level of parity. We can see that the ranking of the four professional sports based on the AUWLC or the AUSWC matches with their ranking based on ASD but not RSD, which does not imply a direct association between the area under the curves and the level of parity. Notice that although NFL has highest ASD, it also has highest ISD since is has the least number of pairwise comparisons. Similarly, the AUWLC and AUSWC of NFL also have greater standard errors when compared with those of the other three sports. Therefore, we can also standardize the AUWLC and the AUSWC by subtracting 0.5 and dividing by $1/\sqrt{G}$. Table~{\ref{table:2}} shows the standardized AUWLC and standardized AUSWC, along with the RSD from Table~{\ref{table:1}}.
\begin{table}
\centering
\begin{tabular}{|c||c|c|c|}
\hline
     League & ASD & ISD & RSD  \\
     \hline
     MLB & 0.071 & 0.039 & 1.812\\ 
     NHL & 0.100 & 0.055 & 1.819\\
     NBA & 0.149 & 0.055 & 2.699\\
     NFL & 0.200 & 0.125 & 1.601\\
     \hline
\end{tabular}
\caption{The actual standard deviation (ASD), idealized standard deviation (ISD), and relative standard deviation (RSD) of 2017(-18) end-of-season win percentages.}
\label{table:1}
\end{table}
\begin{table}
\centering
\begin{tabular}{|c||c|c|c|}
\hline
     League & Standardized AUWLC & Standardized AUSWC & RSD  \\
     \hline
     MLB & 1.387 & 0.865 & 1.812\\ 
     NHL & 1.467 & 0.761 & 1.819\\
     NBA & 2.128 & 1.204 & 2.699\\
     NFL & 1.300 & 0.988 & 1.601\\
     \hline
\end{tabular}
\caption{Comparison of the standardized area under WL-ROC curve (AUWLC), the standardized area under SW-ROC curve (AUSWC), and the relative standard deviation (RSD) of 2017(-18) end-of-season outcomes.}
\label{table:2}
\end{table}

Now the standardized AUWLC provides the same ranking of the level of parity for the four professional sports as the RSD metric. The standardized AUSWC gives a slightly different ranking than the ranking by RSD, which can be expected since the SW-ROC approach is based on the winning percentage of each team when it is predicted to be stronger than the opponent, which is calculated differently from RSD and requires further analysis. Despite that, both approaches show that NBA achieves the least parity among the four professional sports leagues (for similar results and explanations, see \cite{vrooman1995general, rockerbie2016exploring}).

\section{Conclusions}
\label{sec:conclusions}
This paper presents two approaches to extend ROC analysis for paired comparison data that resolve the ambiguity in the coding of paired comparison outcomes. The first approach involves defining sensitivity and specificity relative to a team winning a game, while the second approach involves defining these two measures relative to the team estimated to be stronger within a pair. Each approach can be used to construct ROC curves with area under the curves measuring the discriminating ability between the corresponding ``success'' and ``failure'' outcomes. We demonstrated the application of these approaches using the the regular-season professional sports game outcomes. By standardizing the statistics to account for different numbers of games played in each sports, we conclude that NBA has less parity than the other professional sports.

A few extensions of the current approaches can be considered. For example, in this paper, we only consider the most basic setting where the outcomes are binary, but the approach can be extended to the setting that explicitly incorporates a tie as an outcome. There have been several developments of the paired comparison models to incorporate ties \citep{glenn1960ties, rao1967ties} as well as extensions of ROC analysis to three-class outcomes \citep{mossman1999three}, in which case, an ROC curve becomes an ROC surface, and the area under the curve (AUC) becomes the volume under the surface (VUS).

\appendix

\section{Proof of Theorem~\ref{theorem:convergence}}
\label{app:convergence}
\subsection*{\underline{Convergence of $\widehat{f}^{w\ell}_{\boldsymbol{n}}$ to $f^{w\ell}_{\boldsymbol{n}}$}}

As defined in (\ref{equa:estimatedRwl}) and (\ref{equa:trueRwl}), $\widehat{f}^{w\ell}_{\boldsymbol{n}}$ and $f^{w\ell}_{\boldsymbol{n}}$ are piecewise linear functions with knots $\boldsymbol{\hat{R}^{w\ell}}$ and $\boldsymbol{R^{w\ell}}$ respectively so we first want to show the knots converge. Under the partition assumption by \citet{ford1957solution} and a linear constraint, there exists a unique MLE estimator $\boldsymbol{\hat{\mu}}$ of $\boldsymbol{\mu}$. By the invariance and consistency properties of MLE, $\hat{p}_{ij}$ are consistent estimators of $p_{ij}$. Let \[\epsilon<\min_{r\in[2R-1]}|p_{(r+1)}-p_{(r)}|/2,\]
and let $1\leq z < \infty$ as specified by the definition of the ROC metric (\ref{equa:ROCmetric}),
then there exists $N_1>0$ such that for all nonzero $n_{ij}>N_1$, where $i,j\in[m]$ and $i<j$, 
\begin{equation}
\label{equa:a1}
P(\max_{i,j}|\hat{p}_{ij}-p_{ij}|\geq\frac{\epsilon^{z}}{2^{z+1}R})<\epsilon/2,
\end{equation} 
which means the ranking of $\boldsymbol{\hat{p}}$ with respect to $ij$ and the ranking of $\boldsymbol{p}$ with respect to $ij$ are the same. Since $\boldsymbol{w_{\hat{p}}}$ and $\boldsymbol{w_{p}}$ are ordered according to $\boldsymbol{\hat{p}}$ and $\boldsymbol{p}$, (\ref{equa:a1}) implies that
\begin{equation}
\label{equa:a2}
    P(\boldsymbol{w_{\hat{p}}}=\boldsymbol{w_{p}}) \geq 1-\epsilon/2,
\end{equation}
and since $\boldsymbol{\hat{R}^{w\ell}}$ and $\boldsymbol{R^{w\ell}}$ are functions of $\boldsymbol{w_{\hat{p}}}$ and $\boldsymbol{w_{p}}$, (\ref{equa:a2}) implies that
\begin{equation*}
    P(\boldsymbol{\hat{R}^{w\ell}}=\boldsymbol{R^{w\ell}})\geq 1-\epsilon/2.
\end{equation*}
Assume $\max_{i,j}|\hat{p}_{ij}-p_{ij}|<\frac{\epsilon^{z}}{2^{z+1}R}$. Using the ROC metric (\ref{equa:ROCmetric}), we can calculate the distance between $\widehat{f}_{\boldsymbol{n}}^{w\ell}$ and $f_{\boldsymbol{n}}^{w\ell}$ as
\begin{align*}
    d_{\rho,z}(\widehat{f}_{\boldsymbol{n}}^{w\ell},f_{\boldsymbol{n}}^{w\ell})&=\bigg(\sum_{r=1}^{2R}|\hat{p}_{(r)}-p_{(r)}|\ \rho\Big((0,0),\big(\frac{w_{p_{(2R+r-1)}}}{N},\frac{w_{p_{(r)}}}{N}\big)\Big)^{z}\bigg)^{1/z}\\
    &\leq \Big(\sum_{r=1}^{2R}|\hat{p}_{(r)}-p_{(r)}|\Big)^{1/z}\\
    &\leq \Big(2R\max_{i,j}|\hat{p}_{ij}-p_{ij}|\Big)^{1/z}\\
    &<\epsilon/2.
\end{align*}
Therefore, for all nonzero $n_{ij}>N_1$, where $i,j\in[m]$ and $i<j$,
\begin{equation}
\label{equa:a3}
    P\Big(d_{\rho,z}(\widehat{f}_{\boldsymbol{n}}^{w\ell},f_{\boldsymbol{n}}^{w\ell})\geq \epsilon/2\Big)<\epsilon/2.
\end{equation}

\subsection*{\underline{Convergence of $f^{w\ell}_{\boldsymbol{n}}$ to $\widetilde{f}^{w\ell}_{\boldsymbol{\mathcal{D}}}$}}

The ROC metric between $f_{\boldsymbol{n}}^{w\ell}$ and $\widetilde{f}_{\boldsymbol{\mathcal{D}}}^{w\ell}$ is calculated as
\begin{align*}
   d_{\rho,z}(f_{\boldsymbol{n}}^{w\ell},\widetilde{f}_{\boldsymbol{\mathcal{D}}}^{w\ell})&=\Big(\sum_{k=2}^{2R}(p_{(k)}-p_{(k-1)})\rho_k^{z}\Big)^{1/z}\\
   &\leq (2R)^{1/z}\max_{k\in[2R]}\rho_k,
\end{align*}
where
\begin{align*}
    \rho_k=\rho\Big(&\big(\frac{\sum_{r=1}^{2R-k+1}w_{p_{(r)}}}{N},\frac{\sum_{r=k}^{2R}w_{p_{(r)}}}{N}\big), \\
    &\big(\sum_{r=1}^{2R-k+1}p_{(r)}d_{|r-R-1/2|+1/2},\sum_{r=k}^{2R}p_{(r)}d_{|r-R-1/2|+1/2}\big)\Big).
\end{align*}
By law of large numbers, for $r\in[2R]$, as $n_{p_{(r)}}\to\infty$,
$w_{p_{(r)}}/n_{p_{(r)}}\overset{p}{\longrightarrow} p_{(r)}$.
Assume that $n_{p_{(r)}}/N=d_{|r-R-1/2|+1/2}$ is fixed by design, then as $N\to\infty$,
\[w_{p_{(r)}}/N\overset{p}{\longrightarrow} p_{(r)}d_{|r-R-1/2|+1/2}.\] 
Then the true TPR and FPR converge to the limiting TPR and FPR respectively since for $k\in[2R]$,
\begin{align*}
    \frac{\sum_{r=1}^{2R-k+1}w_{p_{(r)}}}{N}&\overset{p}{\longrightarrow}\sum_{r=1}^{2R-k+1}p_{(r)}d_{|r-R-1/2|+1/2},\\
    \frac{\sum_{r=k}^{2R}w_{p_{(r)}}}{N}&\overset{p}{\longrightarrow}\sum_{r=k}^{2R}p_{(r)}d_{|r-R-1/2|+1/2}.
\end{align*}
The pair of true TPR and FPR also converges to the pair of limiting TPR and FPR
\begin{align*}
    &\Big(\frac{\sum_{r=1}^{2R-k+1}w_{p_{(r)}}}{N},\frac{\sum_{r=k}^{2R}w_{p_{(r)}}}{N}\Big)\overset{p}{\longrightarrow}\\
    &\Big(\sum_{r=1}^{2R-k+1}p_{(r)}d_{|r-R-1/2|+1/2},\sum_{r=k}^{2R}p_{(r)}d_{|r-R-1/2|+1/2}\Big).
\end{align*}
For every metric $\rho$ on $\mathbb{R}^2$ and $1\leq z<\infty$, there exists $N_2>0$ such that for $\{n_{p_{(r)}}:d_{|r-R-1/2|+1/2}\neq0\}>N_2$ as the design $\mathcal{D}$,
\begin{align*}
    P\Big(\max_{k\in[2R]}\rho_k\geq\frac{\epsilon}{2(2R)^{1/z}}\Big)<\epsilon/2,
\end{align*}
and then 
\begin{equation}
\label{equa:a4}
    P\Big(d_{\rho,z}(f_{\boldsymbol{n}}^{w\ell},\widetilde{f}_{\boldsymbol{\mathcal{D}}}^{w\ell})\geq\epsilon/2\Big)
    \leq P\Big((2R)^{1/z}\max_{k\in[2R]}\rho_k\geq\epsilon/2\Big) < \epsilon/2.
\end{equation}

\subsection*{\underline{Convergence of $\widehat{f}^{w\ell}_{\boldsymbol{n}}$ to $\widetilde{f}^{w\ell}_{\boldsymbol{\mathcal{D}}}$}}

Let us combine the results in (\ref{equa:a3}) and (\ref{equa:a4}). Let $N_3=\max\{N_1,N_2\}$,
then for $\{n_{p_{(r)}}:d_{|r-R-1/2|+1/2}\neq0\}>N_3$ as the design $\mathcal{D}$,
\begin{align*}
    P\Big(d_{\rho,z}(\widehat{f}^{w\ell}_{\boldsymbol{n}},\widetilde{f}^{w\ell}_{\boldsymbol{\mathcal{D}}})\geq\epsilon\Big)&\leq P\Big(d_{\rho,z}(\widehat{f}^{w\ell}_{\boldsymbol{n}},f^{w\ell}_{\boldsymbol{n}})\geq\epsilon/2\Big)+P\Big(d_{\rho,z}(f^{w\ell}_{\boldsymbol{n}},\widetilde{f}^{w\ell}_{\boldsymbol{\mathcal{D}}})\geq\epsilon/2\Big)\\
    &<\epsilon/2+\epsilon/2\\
    &=\epsilon.
\end{align*}

\subsection*{\underline{Convergence of $\widehat{f}^{sw}_{\boldsymbol{n}}$ to $f^{sw}_{\boldsymbol{n}}$}}

The proof for SW-ROC curves is similar to the proof for WL-ROC curves shown above. By the same reasoning as (\ref{equa:a1}), for $1\leq z < \infty$ and
\[\epsilon<\min_{r\in[2R-1]}|\hat{q}_{(r+1)}-q_{(r)}|/2,\] there exists $N_4$ such that for all nonzero $n_{ij}>N_4$,
\begin{equation}
\label{equa:a5}
    P(\max_{i,j}|\hat{q}_{ij}-q_{ij}|\geq\frac{\epsilon^{z}}{2^zR})<1-\epsilon/2.
\end{equation}
It follows from (\ref{equa:a5}) that 
\begin{equation*}
    P(\boldsymbol{\hat{R}^{sw}}=\boldsymbol{R^{sw}}) \geq 1-\epsilon/2.
\end{equation*}
Assume $\max_{i,j}|\hat{q}_{ij}-q_{ij}|<\frac{\epsilon_1^{z}}{2^zR}$, then the ROC metric between $\widehat{f}^{sw}_{\boldsymbol{n}}$ and $f^{sw}_{\boldsymbol{n}}$ is calculated as
\begin{align*}
    d_{\rho,z}(\widehat{f}_{\boldsymbol{n}}^{sw},f_{\boldsymbol{n}}^{sw})&=\bigg(\sum_{r=1}^{R}|\hat{q}_{(r)}-q_{(r)}|\ \rho\Big((0,0),\big(\frac{n_{q_{(r)}}-w_{q_{(r)}}}{N-W},\frac{w_{q_{(r)}}}{W}\big)\Big)^{z}\bigg)^{1/z}\\
    &\leq \Big(\sum_{r=1}^{R}|\hat{q}_{(r)}-q_{(r)}|\Big)^{1/z}\\
    &\leq \Big(R\max_{i,j}|\hat{q}_{ij}-q_{ij}|\Big)^{1/z}\\
    &<\epsilon/2.
\end{align*}
Therefore, for all nonzero $n_{ij}>N_4$, where $i,j\in[m]$ and $i<j$,
\begin{equation}
\label{equa:a6}
    P\Big(d_{\rho,z}(\widehat{f}_{\boldsymbol{n}}^{sw},f_{\boldsymbol{n}}^{sw})\geq \epsilon/2\Big)< \epsilon/2.
\end{equation}

\subsection*{\underline{Convergence of $f^{sw}_{\boldsymbol{n}}$ to $\widetilde{f}^{sw}_{\boldsymbol{\mathcal{D}}}$}}

The ROC metric between $f_{\boldsymbol{n}}^{sw}$ and $\widetilde{f}_{\boldsymbol{\mathcal{D}}}^{sw}$ can be calculated as
\begin{align*}
   d_{\rho,z}(f_{\boldsymbol{n}}^{sw},\widetilde{f}_{\boldsymbol{\mathcal{D}}}^{sw})&=\Big(\sum_{k=2}^{R}(q_{(k)}-q_{(k-1)})\rho_k^{z}\Big)^{1/z}\\
   &\leq R^{1/z}\max_{k\in[R]}\rho_k,
\end{align*}
where
\[\rho_k=\rho\Big(\big(\frac{\sum_{r=k}^{R}n_{q_{r}}-w_{q_{(r)}}}{N-W},\frac{\sum_{r=k}^{R}w_{q_{(r)}}}{W}\big),\big(\frac{\sum_{r=k}^{R}d_{r}(1-q_{(r)})}{\sum_{r=1}^{R}d_{r}(1-q_{(r)})},\frac{\sum_{r=k}^{R}d_{r}q_{(r)}}{\sum_{r=1}^{R}d_{r}q_{(r)}}\big)\Big).\]
For $r\in[R]$, as $n_{q_{(r)}}\to\infty$,
$w_{q_{(r)}}/n_{q_{(r)}}\overset{p}{\longrightarrow} q_{(r)}$. If $N\to\infty$ with fixed ratio $n_{q_{(r)}}/N=d_{r}$, then
$w_{q_{(r)}}/N\overset{p}{\longrightarrow} d_{r}q_{(r)}$ and $W/N\overset{p}{\longrightarrow} \sum_{r=1}^{R}d_{r}q_{(r)}$. 
The pairs of TPR and FPR also converge 
\begin{align*}
    \Big(\frac{\sum_{r=k}^{R}w_{q_{(r)}}}{W},\frac{\sum_{r=k}^{R}n_{q_{(r)}}-w_{q_{(r)}}}{N-W}\Big)\overset{p}{\longrightarrow} \Big(\frac{\sum_{r=k}^{R}d_{r}q_{(r)}}{\sum_{r=1}^{R}d_{r}q_{(r)}},\frac{\sum_{r=k}^{R}d_{r}(1-q_{(r)})}{\sum_{r=1}^{R}(1-d_{r}q_{(r)})}\Big).
\end{align*}
For every metric $\rho$ on $\mathbb{R}^2$ and $1\leq z <\infty$, there exists $N_5>0$ such that for $\{n_{q_{(r)}}:d_{r}\neq0\}>N_5$,
\begin{align*}
    P\Big(\max_{k\in[R]}\rho_k\geq\frac{\epsilon}{2R^{1/z}}\Big)<\epsilon/2,
\end{align*}
and then 
\begin{equation}
\label{equa:a7}
    P(d_{\rho,z}(f_{\boldsymbol{n}}^{sw},\widetilde{f}_{\boldsymbol{\mathcal{D}}}^{sw})\geq\epsilon/2)
    \leq P\Big(R^{1/z}\max_{k\in[R]}\rho_k\geq\epsilon/2\Big) < \epsilon/2.
\end{equation}

\subsection*{\underline{Convergence of $\widehat{f}^{sw}_{\boldsymbol{n}}$ to $\widetilde{f}^{sw}_{\boldsymbol{\mathcal{D}}}$}}

Let $N_6=\max\{N_5,N_6\}$. It follows from (\ref{equa:a6}) and (\ref{equa:a7}) that for $\{n_{q_{(r)}}:d_{r}\neq0\}>N_6$, 
\begin{align*}
    P(d_{\rho,z}(\widehat{f}^{sw}_{\boldsymbol{n}},\widetilde{f}^{sw}_{\boldsymbol{\mathcal{D}}})\geq\epsilon)&\leq P(d_{\rho,z}(\widehat{f}^{sw}_{\boldsymbol{n}},f^{sw}_{\boldsymbol{n}})\geq\epsilon/2)+P(d_{\rho,z}(f^{sw}_{\boldsymbol{n}},\widetilde{f}^{sw}_{\boldsymbol{\mathcal{D}}})\geq\epsilon/2)\\
    &<\epsilon/2+\epsilon/2\\
    &=\epsilon.
\end{align*}

\subsection*{\underline{Final Step}}
Let $N^*=\max\{N_3,N_6\}$, and then the proof of Theorem~\ref{theorem:convergence} is complete.

\section{Proofs in Section~\ref{sec:comparison}}
\subsection{Proof of the linear relationship between \texorpdfstring{$\hat{c}^{w\ell}$}{c-wl} and \texorpdfstring{$\hat{c}^{sw}$}{c-sw} (\ref{equa:linear})}
\label{app:linear}

As defined in (\ref{equa:estimatedCwl}), $\hat{c}^{w\ell}$ is calculated as
\[\hat{c}^{w\ell}=\frac{\sum_{a,b}\mathbb{1}_{\{\hat{p}^{w}_a>\hat{p}^{\ell}_b\}}+\frac{1}{2}\mathbb{1}_{\{\hat{p}^{w}_a=\hat{p}^{\ell}_b\}}}{N^2}.\]
Thus,
\begin{align}
\label{equa:b1}
    N^2\hat{c}^{w\ell}
    =&\sum_{r=1}^{R}\Big((n_{\hat{q}_{(r)}}-w_{\hat{q}_{(r)}})\sum_{s=r+1}^{R}w_{\hat{q}_{(s)}}\Big)+\sum_{r=1}^{R}w_{\hat{q}_{(r)}}(n_{\hat{q}_{(r)}}-w_{\hat{q}_{(r)}}) \nonumber \\
    &+\sum_{r=1}^{R}\Big(w_{\hat{q}_{(r)}}\big(\sum_{s=1}^{R}w_{\hat{q}_{(s)}}+\sum_{s=1}^{r-1}(n_{\hat{q}_{(s)}}-w_{\hat{q}_{(s)}})\big)\Big) \nonumber \\ 
    =& 2\sum_{r=1}^{R}\Big(w_{\hat{q}_{(r)}}\sum_{s=1}^{r-1}n_{\hat{q}_{(s)}}\Big)-2\sum_{r=1}^{R}\Big(w_{\hat{q}_{(r)}}\sum_{s=1}^{r-1}w_{\hat{q}_{(s)}}\Big)+\Big(\sum_{r=1}^{R}w_{\hat{q}_{(r)}}\Big)^2 \nonumber \\
    &+\sum_{r=1}^{R}n_{\hat{q}_{(r)}}w_{\hat{q}_{(r)}}-\sum_{r=1}^{R}w_{\hat{q}_{(r)}}^2 .
\end{align}
Since
$\Big(\sum_{r=1}^{R}w_{\hat{q}_{(r)}}\Big)^2=\sum_{r=1}^{R}w_{\hat{q}_{(r)}}^2+2\sum_{r=1}^{R}w_{\hat{q}_{(r)}}\sum_{s=1}^{r-1}w_{\hat{q}_{(s)}}$, (\ref{equa:b1}) implies that
\begin{align}
\label{equa:b2}
    N^2\hat{c}^{w\ell}=&2\sum_{r=1}^{R}\Big(w_{\hat{q}_{(r)}}\sum_{s=1}^{r-1}n_{\hat{q}_{(s)}}\Big)+\sum_{r=1}^{R}n_{\hat{q}_{(r)}}w_{\hat{q}_{(r)}} =2\sum_{r=1}^{R}\hat{A}_rw_{\hat{q}_{(r)}},
\end{align}
where $\hat{A}_r=\sum_{s=1}^{r-1}n_{\hat{q}_{(s)}}+\frac{1}{2}n_{\hat{q}_{(r)}}$. Similarly, as shown in (\ref{equa:estimatedCsw}), $\hat{c}^{sw}$ is calculated as
$$\hat{c}^{sw}=\frac{\sum_{r,s}\mathbb{1}_{\{\hat{p}^{st}_r>\hat{p}^{wk}_s\}}+\frac{1}{2}\mathbb{1}_{\{\hat{p}^{st}_r=\hat{p}^{wk}_s\}}}{\hat{W}(N-\hat{W})},$$
then
\begin{align}
\label{equa:b3}
    \hat{W}(N-\hat{W})\hat{c}^{sw} &=\sum_{r=1}^{R}\Big(w_{\hat{q}_{(r)}}\sum_{s=1}^{r-1}(n_{\hat{q}_{(s)}}-w_{\hat{q}_{(s)}})\Big)+\frac{1}{2}\sum_{r=1}^{R}w_{\hat{q}_{(r)}}(n_{\hat{q}_{(r)}}-w_{\hat{q}_{(r)}}) \nonumber \\
    &= \sum_{r=1}^{R}\Big(w_{\hat{q}_{(r)}}\sum_{s=1}^{r-1}n_{\hat{q}_{(s)}}\Big)-\frac{1}{2}\Big(\sum_{r=1}^{R}w_{\hat{q}_{(r)}}\Big)^2+\frac{1}{2}\sum_{r=1}^{R}n_{\hat{q}_{(r)}}w_{\hat{q}_{(r)}} \nonumber \\
    &=\sum_{r=1}^{R}\hat{A}_rw_{\hat{q}_{(r)}}-\frac{1}{2}(\hat{W})^2.
\end{align}
Therefore, the relationship between $\hat{c}^{w\ell}$ and $\hat{c}^{sw}$ can be established from (\ref{equa:b2}) and (\ref{equa:b3}) that
\begin{align*}
    N^2\hat{c}^{w\ell}=2\hat{W}(N-\hat{W})\hat{c}^{sw}+(\hat{W})^2.
\end{align*}

\subsection{Proof of \texorpdfstring{$\hat{c}^{w\ell}\geq\hat{c}^{sw}$}{c-wl>c-sw} (\ref{equa:c-compare})}
\label{app:c-compare}
The limiting WL-c-statistic $\tilde{c}^{w\ell}$ and the limiting SW-c-statistic $\tilde{c}^{sw}$ are defined in (\ref{equa:limitinngCwl}) and (\ref{equa:limitinngCsw}) as
\begin{align*}
    \tilde{c}^{w\ell}
    &=\frac{2\sum_{r=1}^{R}(r-\frac{1}{2})q_{(r)}}{R^2},\\
    \tilde{c}^{sw}
    &= \frac{\sum_{r=1}^{R}(r-\frac{1}{2})q_{(r)}-\frac{1}{2}\big(\sum_{r=1}^{R}q_{(r)}\big)^2}{\big(\sum_{r=1}^{R}q_{(r)}\big)\big(R-\sum_{r=1}^{R}q_{(r)}\big)}.
\end{align*}
Let 
\begin{equation*}
    U = \sum_{r=1}^{R}(r-\frac{1}{2})q_{(r)},\ V = \sum_{r=1}^{R}q_{(r)}, \text{ and }
    \bar{V} = \frac{V}{R},
\end{equation*}
then
\begin{align*}
    \tilde{c}^{w\ell}&=\frac{2U}{R^2},\\
    \tilde{c}^{sw}&=\frac{U-\frac{1}{2}V^2}{V(R-V)}.
\end{align*}
Therefore, $\hat{c}^{w\ell}\geq \hat{c}^{sw}$ is equivalent to
\begin{equation}
\label{equa:b4}
    U\leq\frac{\bar{V}^2R^2}{2-4\bar{V}(1-\bar{V})}.
\end{equation}
To prove $\hat{c}^{w\ell}\geq \hat{c}^{sw}$, we just need to prove the ineqality in (\ref{equa:b4}) holds.

Let us first consider the case when there are only 2 teams, that is $m=2$ and $R=1$, then $2U=V=\bar{V}=q_{(1)}\in[0.5,1)$. Since \[U\leq\frac{\bar{V}^2R^2}{2-4\bar{V}(1-\bar{V})}=\frac{2U^2}{1-4U+8U^2}\] for $q_{(1)}\in[0.5,1)$, (\ref{equa:b4}) holds when $m=2$. 

Now let us assume $m=3$ and $R=3$. For a given $\bar{V}$, if we want to maximize $U$, we should try to minimize $q_{(1)}$ and maximize $q_{(3)}$. We know that $q_{(1)}$, $q_{(2)}$, and $q_{(3)}$ are all between $0.5$ and $1$. To set an upper bound for $U$, we let $q_{(1)}=q_{(2)}=0.5$ and $q_{(3)}=3\bar{V}-1$ when $\bar{V}\in[1/2, 2/3)$; let $q_{(1)}=0.5$, $q_{(2)}=3\bar{V}-1.5$, and $q_{(3)}=1$ when $\bar{V}\in[2/3, 5/6)$; and let $q_{(1)}=3\bar{V}-2$ and $q_{(2)}=q_{(3)}=1$ when $\bar{V}\in[5/6, 1)$. That is, $U$ is upper bounded by
$$U\leq\left\{
  \begin{array}{@{}ll@{}}
    7.5\bar{V}-1.5,& \bar{V}\in[1/2,2/3),\\
    4.5\bar{V}+0.5,& \bar{V}\in[2/3,5/6),\\
    1.5\bar{V}+3,&  \bar{V}\in[5/6,1).
  \end{array}\right.$$
\begin{figure}[ht]
    \centering
    \includegraphics[scale=0.5]{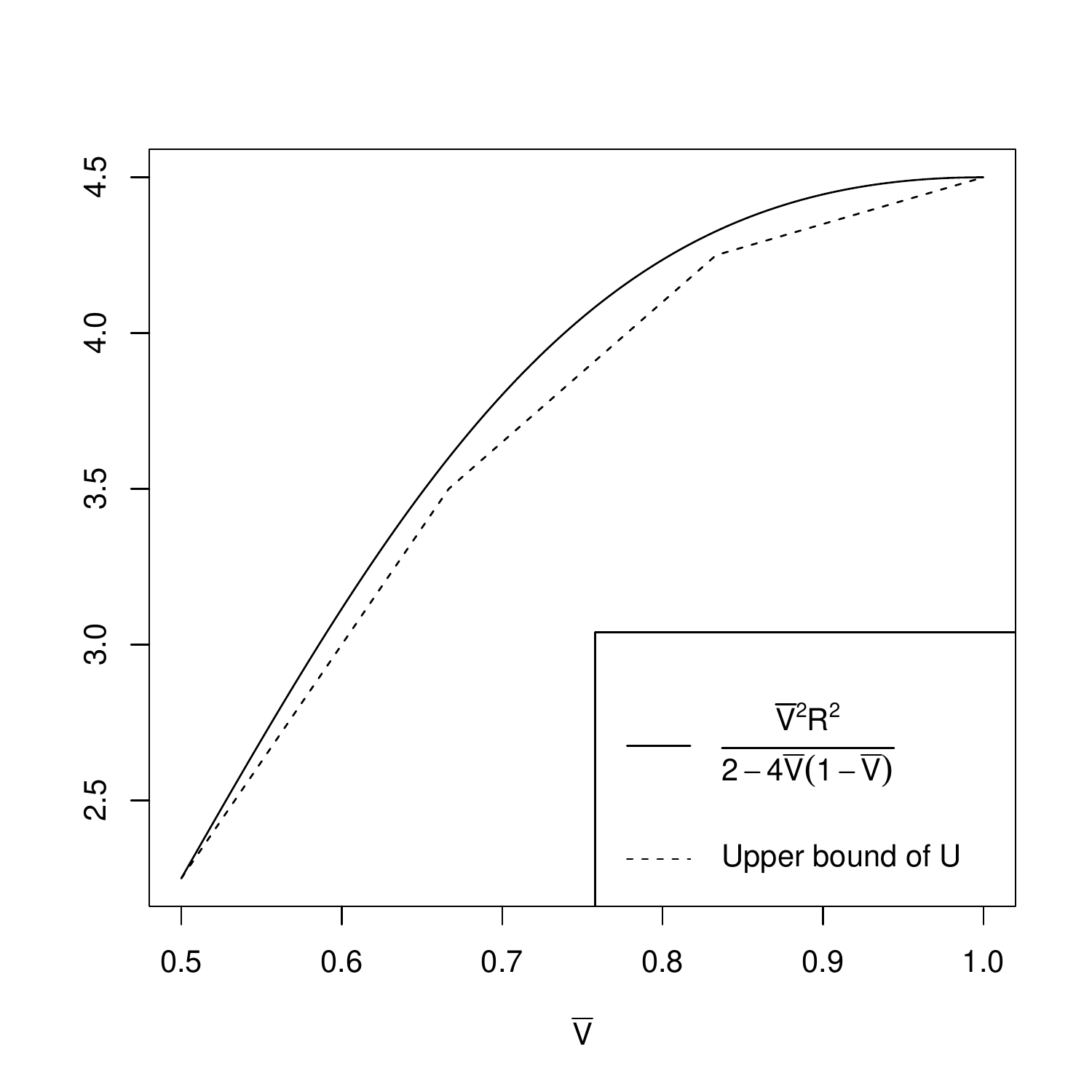}
    \caption{Plot of both sides of (\ref{equa:b4}) versus $\bar{V}$ when there are 3 teams.}
    \label{fig:m=3}
\end{figure}
We just need to check if (\ref{equa:b4}) is satisfied for $\bar{V}\in[0.5,1)$. In Figure~\ref{fig:m=3}, we plotted the upper bound of $U$ along with the right-hand side of (\ref{equa:b4}), which is $$\Big\{\frac{9\bar{V}^2}{2-4\bar{V}(1-\bar{V})}:\bar{V}\in[0.5,1)\Big\}.$$ Clearly, the condition is satisfied for $\bar{V}\in[0.5,1)$. Hence (\ref{equa:b4}) holds when $m=3$.

In general, when there are $m$ teams and $R=m(m-1)/2$ pairings, the idea is to construct an upper bound for $U$ and to show that the upper bound is always less than the right-hand side of (\ref{equa:b4}). If $\bar{V}\in[\frac{R+k-1}{2R},\frac{R+k}{2R})$, $k=1,\dots,R$, then $U$ is upper bounded by setting $p_{(R+1)}=\dots=p_{(2R-k)}=0.5$, $p_{(2R-k+1)}=(\bar{V}-0.5)R-0.5k+1$, and $p_{(2R-k+2)}=\dots=p_{(2R)}=1$. That is, for $\bar{V}\in[\frac{R+k-1}{2R},\frac{R+k}{2R})$, 
$$U\leq R(R-k+0.5)\bar{V}-0.25(R-k)(R-k+1)+0.5k(k-1).$$ Hence the upper bound we constructed for $U$ is a piece-wise linear function of $\bar{V}\in[0.5,1)$, denoted as $u(\bar{V})$. Let $g(\bar{V})$ be a function of $\bar{V}$ defined on $[0.5,1)$ that equals to the right hand side of (\ref{equa:b4}). Then we just need to prove that $$u(\bar{V})\leq g(\bar{V}), \bar{V}\in[0.5,1).$$
Since $g(\bar{V})$ is a concave function of $\bar{V}$, it suffices to prove that each knot of $u(\bar{V})$ lies on or below $g(\bar{V})$.
For $k=0,\dots,R$,
\begin{align*}
  u(\frac{R+k}{2R})&=0.25(R^2+2kR-k^2)\\
  g(\frac{R+k}{2R})&=0.25(\frac{2kR^3}{R^2+k^2}+R^2)
\end{align*}
We have
$$g(0.5+\frac{k}{2R})-u(0.5+\frac{k}{2R})=\frac{0.25k^2(R-k)^2}{R^2+k^2}\geq 0.$$
Therefore, (\ref{equa:b4}) holds for $m\geq2$, and the proof is complete.



\bibliographystyle{imsart-nameyear} 
\bibliography{reference.bib}       


\end{document}